\documentclass[aps,pre,superscriptaddress,showkeys,twocolumn,longbibliography]{revtex4-2}

\usepackage{amsmath}
\usepackage{graphicx}
\usepackage{float,physics}
\usepackage{siunitx}
\usepackage{xcolor}
\usepackage{epstopdf}
\definecolor{darkgreen}{rgb}{0,0.6,0.0}

\usepackage{latexsym}
\usepackage{amsmath}
\usepackage{amsfonts}
\usepackage{amssymb}
\usepackage{mathrsfs}
\usepackage{verbatim}
\usepackage{anysize}
\usepackage{soul}

\usepackage[utf8]{inputenc}
\usepackage{algorithmic}
\usepackage{algorithm}
\usepackage{enumerate}
\usepackage{amssymb, amsmath, amsthm}
\usepackage{graphicx}
\usepackage{lmodern,url}
\usepackage{graphicx}
\usepackage{wrapfig}

\floatname{algorithm}{Algoritmo}

\newcommand{\bo}{\raise-1mm\hbox{\Large$\Box$}}

\begin{document}

\title{Wetting dynamics by mixtures of fast and slow self-propelled particles}
\normalsize
\author{Mauricio Rojas-Vega}

\affiliation{Departamento de F\'isica, FCFM, Universidad de Chile, Santiago, Chile}
\affiliation{Institute of Science and Technology Austria, Klosterneuburg, Austria}

\author{Pablo de Castro}

\affiliation{ICTP South American Institute for Fundamental Research \& Instituto de F\'isica Te\'orica, Universidade Estadual Paulista - UNESP, S\~ao Paulo, Brazil}

\author{Rodrigo Soto}

\affiliation{Departamento de F\'isica, FCFM, Universidad de Chile, Santiago, Chile}

\date{\today}

\begin{abstract}
We study active surface wetting using a minimal model of bacteria that takes into account the intrinsic motility diversity of living matter. A mixture of ``fast'' and ``slow'' self-propelled Brownian particles is considered in the presence of a wall. The evolution of the wetting layer thickness shows an overshoot before stationarity and its composition evolves in two stages, equilibrating after a slow elimination of excess particles. Non-monotonic evolutions are shown to arise from delayed avalanches towards the dilute phase combined with the emergence of a transient particle front.
\end{abstract}

\maketitle

\section{Introduction}
Natural active matter, such as collections of organisms, is \textit{not} composed of identical self-propelling agents \cite{peruani2012collective}. Instead, a wide distribution of motility properties exists due to different ages, reproduction stages, shapes, and sizes \cite{berdakin2013quantifying,ipina2019bacteria,berg2008coli}. Moreover, active particles typically interact with ``surfaces'', e.g., bacteria swimming near boundaries of their host body or of contaminated medical instruments \cite{lopez2014dynamics,satpathy2016review}. For simplicity, models usually ignore at least one of these two ingredients, i.e., diversity and surface effects. 

A persistent particle has a self-propulsion direction that fluctuates stochastically and, typically, slowly \cite{andrea2020}. Consequently, active matter accumulates on surfaces to an extent dependent on persistence and density \cite{sepulveda2017wetting}. 
For bacteria, this mechanism, together with other factors, contributes to initiate biofilm formation~\cite{grobas2021swarming}. 
Surface accumulation by persistence is called active wetting \cite{wittmann2016active,turci2021wetting,neta2021wetting,sepulveda2018universality}. Three phases are possible \cite{sepulveda2017wetting}: complete wetting, where the wetting layer covers the wall completely; incomplete wetting, where only a fraction of the wall becomes covered; and ``unwetting'' or ``drying'', where no dense phase exists. 
Active wetting was studied mostly for identical particles. However, passive and active phase behaviors can depend strongly on ``diversity'' in some particle attribute \cite{de2021active,kumar2021effect,PabloPeter1,stenhammar2015activity,de2021diversity, kolb2020active,hoell2019multi,wittkowski2017nonequilibrium,PabloPeter3,takatori2015theory,curatolo2020cooperative,van2020predicting,dolai2018phase,schmid2001wetting,PabloPeter2,williams2022confinement}.

In this Letter, we study a mixture of ``fast'' and ``slow'' active Brownian disks moving in 2D, in the presence of a flat impenetrable wall. Each type has its own self-propulsion speed, defining a degree of \textit{speed diversity}. Besides simulations, a dynamical kinetic theory is developed by extending the approach of Redner \textit{et al.}~\cite{redner} in three fronts: to mixtures, to include walls, and to incorporate time-dependence. This approach calculates the absorption and emission rates for the agglomerate directly from microscopic considerations and is therefore different than free-energy-like approximations \cite{wittmann2017effective} or phenomenological theories \cite{wittkowski2014scalar} that can be harder to connect with microscopic properties. Our theory relies on one fitting parameter only (similarly to Redner's original theory \cite{redner}), which assumes a single value across all parameters, somewhat like a ``universal constant''. To isolate surface effects, we choose a range of densities that allows for significant complete wetting while bulk motility-induced phase separation (MIPS) remains absent. Instead of focusing on ``equilibrium'' wetting-drying transitions \cite{sepulveda2017wetting}, we study the wetting dynamics, i.e., the mechanisms involved in setting the composition and thickness of the wetting layer versus time and how motility diversity affects those. A two-stage evolution is found. Moreover, we identify a transient overshoot of the layer thickness, which occurs even without diversity but whose intensity depends non-monotonically on it. 

\section{Model and simulation method}
We consider a binary mixture in 2D composed of $N$ active Brownian disks (labeled by $i$) where $N/2$ of them are ``fast'' particles, with self-propulsion speed ${v_i=v_\text{f}\equiv v_0 (1+\delta)}$, and the other $N/2$ are ``slow'' particles, with ${v_i=v_\text{s}\equiv v_0(1-\delta)}$. Thus $\delta \in [0,1]$ is the degree of speed diversity \footnote{In other self-clustering problems, binary mixtures were shown to behave similarly to fully polydisperse systems \cite{sollich2001predicting,de2021active,de2021diversity,PabloPeter1,PabloPeter2,PabloPeter3}. Changing our $\delta$ is a proxy for changing the standard deviation of a continuous distribution of speeds.}.
Hereafter ``f'' and ``s'' denote ``fast'' and ``slow'' particles, respectively. Their dynamics obeys
\begin{equation}
    \partial_t\boldsymbol{r}_i=v_{i}\,\hat{\boldsymbol{\nu}}_i+\mu \boldsymbol{F}_i+\boldsymbol{\xi}_i,\quad \partial_t\theta_i=\eta_i (t),
\end{equation}
where ${\hat{\boldsymbol{\nu}}_i=(\cos\theta_i, \sin\theta_i)}$ is the self-propulsion direction, $\mu$ is the mobility and ${\boldsymbol{F}_i=\sum_{j} \boldsymbol{F}_{ij}+\boldsymbol{F}^{\text{wall}}_i}$ is the net force on particle $i$ due to interactions with other particles and with the wall. The noise terms $\boldsymbol{\xi}_i (t)$ and $\eta_i (t)$ are Gaussian and white, with zero mean and correlations ${\langle \xi_{i\lambda}(t) \xi_{j\beta} (t')\rangle =2\xi \delta_{ij}\delta_{\lambda \beta}\delta(t-t')}$ (the Greek letters denote coordinates) and ${\langle \eta_i(t) \eta_j (t')\rangle =2\eta \delta_{ij}\delta(t-t')}$, where $\xi$ and $\eta$ are the translational \footnote{Translational diffusion is included to facilitate (future) theoretical developments and comparisons but it does not affect the qualitative behavior.} and rotational diffusion coefficients, respectively. 

We model particle interactions by a soft repulsive WCA-like potential \cite{maloney2020clustering},
$U=2^{3/2}(\sigma_{ij}/r_{ij})^3 - 3(\sigma_{ij}/r_{ij})^6 + (\sigma_{ij}/r_{ij})^{12}-3/4$
for $r_{ij} \le  2^{\frac{1}{6}} \sigma_{ij}$ and $U=0$ otherwise \footnote{Also to facilitate (future) theoretical developments, the modified WCA potential used here has a smooth second derivative.}, with $r_{ij}$ the interparticle distance and ${\sigma_{ij}\equiv\frac{1}{2}(d_i+d_j)}$, where $d_i$ is the diameter of particle $i$. To avoid crystallization \cite{desmond2009random}, each particle is randomly assigned one of two diameters, $d_\text{small}=d_0$ or $d_\text{large}=1.4d_0$, uncorrelated with self-propulsion speeds. We focus on speed diversity effects and thus the system is said to be just binary (the observed size segregation is weak). We choose $v_0=1$, $d_0=1$, $\mu=1$, $\xi=5\times 10^{-4}$, $\eta=5\times 10^{-3}$ and the forward Euler method with time step $\Delta t= 10^{-4}$. Initially, positions and velocity directions are randomly distributed independent of types. 

\begin{figure}[!h]
	\includegraphics[width=.42\textwidth]{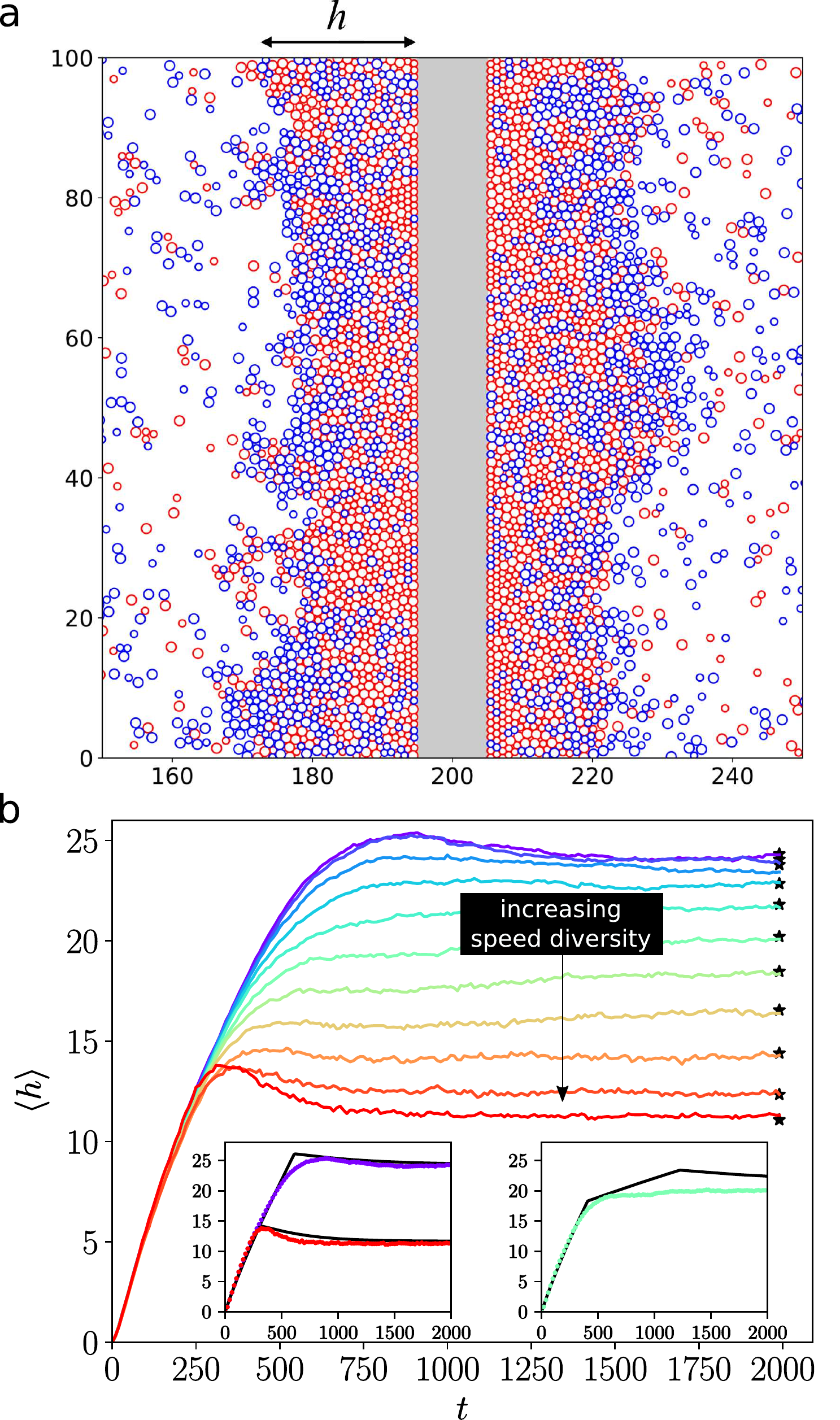}
	\caption{(a) Snapshot for speed diversity $\delta=0.5$ and $\phi=0.18$ in the SS. Fast (slow) particles are in red (blue). 
		(b) Temporal evolution of the mean wetting layer thickness for $\phi=0.18$ and $\delta$ from $0$ to $1$ every $0.1$. Stars indicate SS averages ($t > 100\tau$). Insets: $\delta=0$ and $\delta=1$ on the left and $\delta=0.5$ on the right. Solid lines are the theory.}
	\label{init}
\end{figure}

Figure~\ref{init}(a) shows the system in the steady state (SS). The simulation box---which has total dimensions $L_x=400$ and $L_y=100$ and periodic boundary conditions in the $y$ direction---is shown only partially. An impenetrable flat wall (with sides at $x=195$ and $x=205$) is placed at the center. For particle-wall interactions, the same potential is used with $d_j=0$. The occupied area fraction $\phi$ is the total area occupied by particles divided by the area of the simulation box minus the wall. In all simulations, $\phi=0.18$, i.e., $6000$ particles, leading to complete wetting during the whole dynamics without bulk MIPS, and we focus on varying $\delta$. The average free-particle persistence length is $\ell \equiv v_0/\eta=200$, which is comparable to the system size but sufficiently small to avoid ballistic motion between wall sides. Thus, each wall is treated independently and we average data from both sides. 
For complete wetting, increasing $\phi$ or $\ell$ trivially increases the layer thickness.
For SS averages, only configurations after $t = 100\tau$ were used, where ${\tau\equiv\eta^{-1}=200}$ is the rotational diffusion time.

\section{Wetting}
To characterize accumulation, two particles were considered ``connected'' if ${r_{ij}<1.1\sigma_{ij}}$, allowing us to identify the cluster of connected particles in contact with each wall. The mean wetting layer thickness $\langle h\rangle$ is obtained by averaging the position of the outermost particle in each of the 128 bins in which $L_y$ is divided. Figure~\ref{init}(b) shows $\langle h\rangle(t)$.  The initial growth rate is constant and independent of $\delta$. However, at long times, the higher the $\delta$, the thinner the SS layer. For $\delta=1$ (active-passive mixture), the thickness is approximately half the value for $\delta=0$ as passive disks cannot wet. The layer thickness exhibits a transient overshoot before reaching stationarity. The gap between the peak of $\langle h\rangle(t)$ and $\langle h\rangle(t\rightarrow\infty)$ depends on $\delta$. Such overshoot will be elucidated below.

\begin{figure}[htbp]
    \includegraphics[width=.5\textwidth]{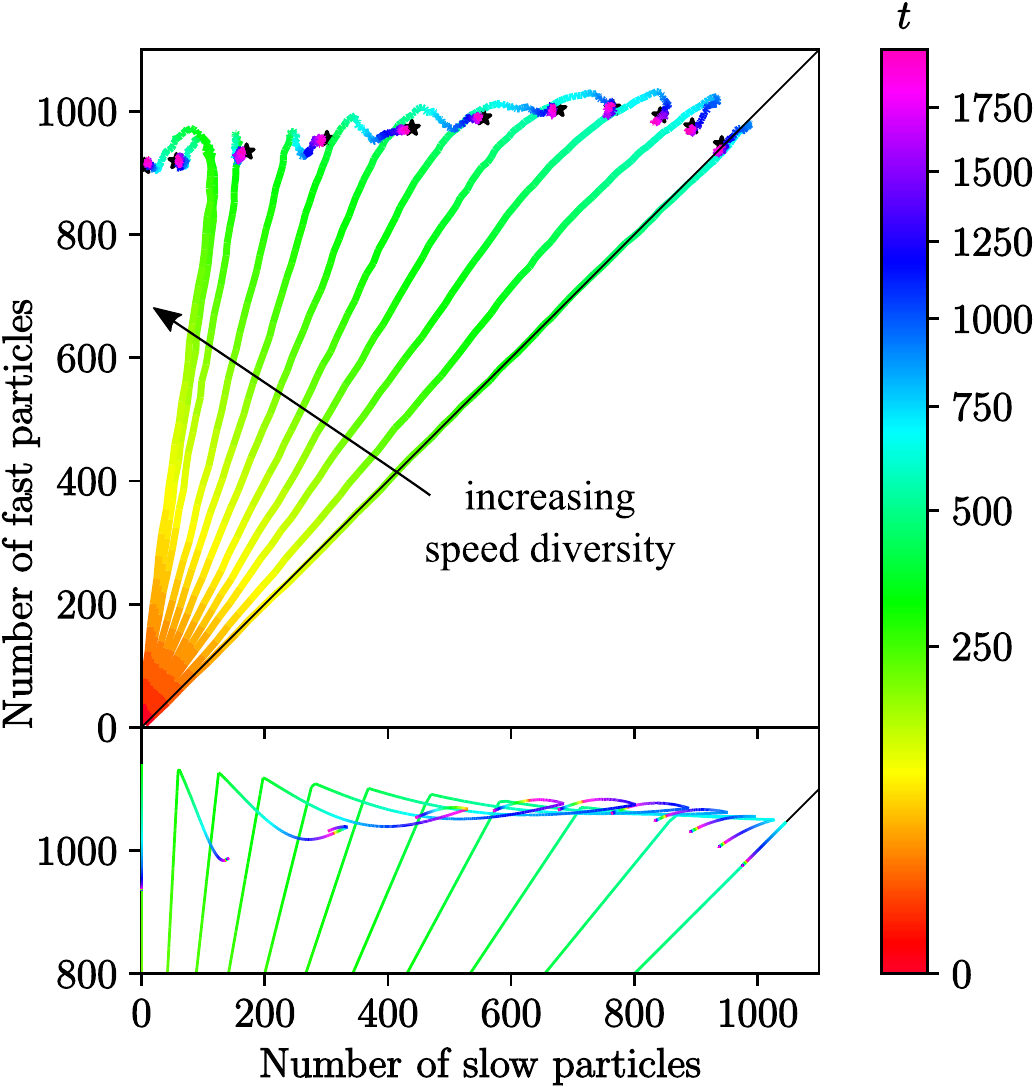}
    \caption{Top: Wetting layer composition evolution for $\delta$ from $0$ to $1$ every $0.1$ and $\phi=0.18$ ($6000$ particles in total), measured by the amount of fast and slow particles in it. The diagonal corresponds to the same amount of fast and slow particles. Star symbols show results for the SS. The colorbar shows time (in a nonlinear scale, to focus on the final stage). Bottom: Theory.}
    \label{Composition}
\end{figure}

The evolution of the layer composition is shown in Figure~\ref{Composition} (see Movie at \href{https://bit.ly/3GgIuBN}{bit.ly/3GgIuBN}). The layer is always richer in fast particles than the overall system. In phase-separation problems, this is known as ``fractionation'' \cite{PabloPeter1,de2021diversity}. The fractionation degree, however, is not constant. There is a first stage where the ratio of fast and slow particles remains constant, depending on $\delta$. A second slower stage then starts, in which the composition is finely adjusted towards the SS. For $\delta>0.7$, excess slow particles are eliminated, while for $\delta<0.7$, additional slow particles are incorporated. This slow dynamics occurs simultaneously with changes in $\langle h\rangle$, with both processes being non-monotonic.

\begin{figure}[htbp]
	\centering
	\includegraphics[width=.5\textwidth]{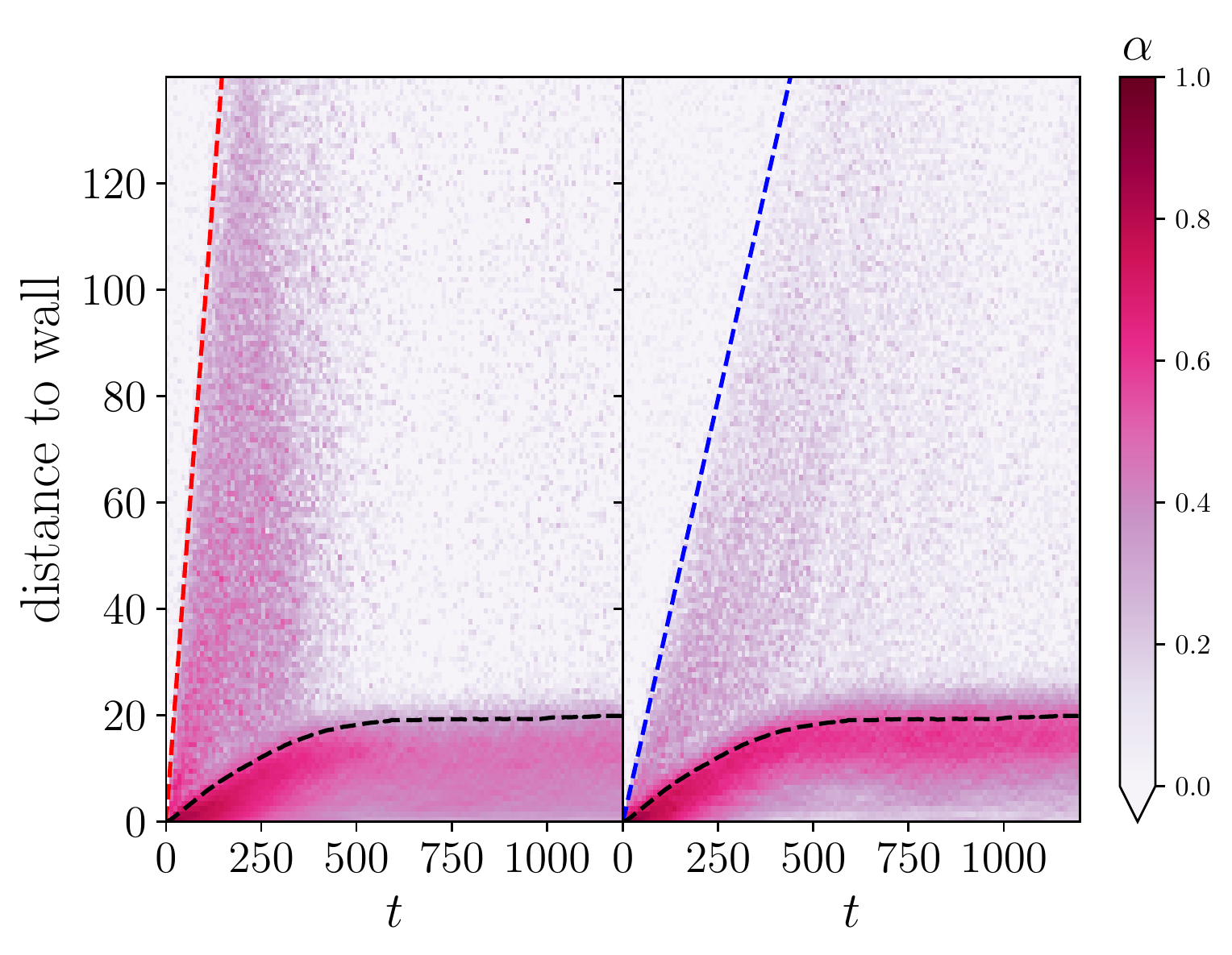}
	\caption{Spatiotemporal diagram of the self-propulsion orientation parameter $\alpha(x)= \langle\hat{n}\cdot \hat{\nu}\rangle$, for fast (left) and slow (right) particles, with $\delta=0.5$ and $\phi=0.18$. The black dashed lines show the mean thickness data shown in Fig.~\ref{init}(b). The dashed red and blue lines are the ``cleaning signals''  with velocities $\langle v_{x}^{(\rm f/s)}\rangle$ (see main text).}
	\label{orievol}
\end{figure}

To model the accumulation dynamics, we first analyze the spatial distribution of orientations by computing $\alpha(x)\equiv\langle\hat{n}\cdot\hat{\nu}\rangle$, shown in Fig.\ \ref{orievol},  where $\hat{n}$ is the inwards normal to the walls and the average is performed over all particles within a $y$-axis stripe of width $d_0$, centered at position $x$. 
Initially, all particles are randomly oriented, implying ${\alpha(x)=0}$ everywhere. Later, particles pointing away from the walls abandon them, leaving regions close to the walls with particles mostly moving towards them. This manifests as regions of ${\alpha(x)>0}$ which grow linearly in time; see red and blue lines. This ``cleaning signal''  has the mean velocity in the $x$ direction at which a randomly oriented particle joins a wall, $\langle v_{x}^{(\rm f/s)}\rangle=\int_{-\pi/2}^{\pi/2}v_{\rm f/s}\cos \theta d\theta/\pi=\frac{2v_{\rm f/s}}{\pi}$, considering only particles moving towards the wall.
At the interface, particles must point towards the wall as otherwise they escape. Consequently, the maximum of $\alpha(x)$ independently locates the interface [see Fig.~\ref{orievol} and compare with the thickness from Fig.~\ref{init}(b)]. This maximum in polarization near the interface, pointing toward the layer, is consistent with simulations and theoretical predictions for ABPs \cite{hermann2020active, paliwal2018chemical, hermann2019phase, solon2018generalized}.

\begin{figure}[htbp]
\includegraphics[width=.9\columnwidth]{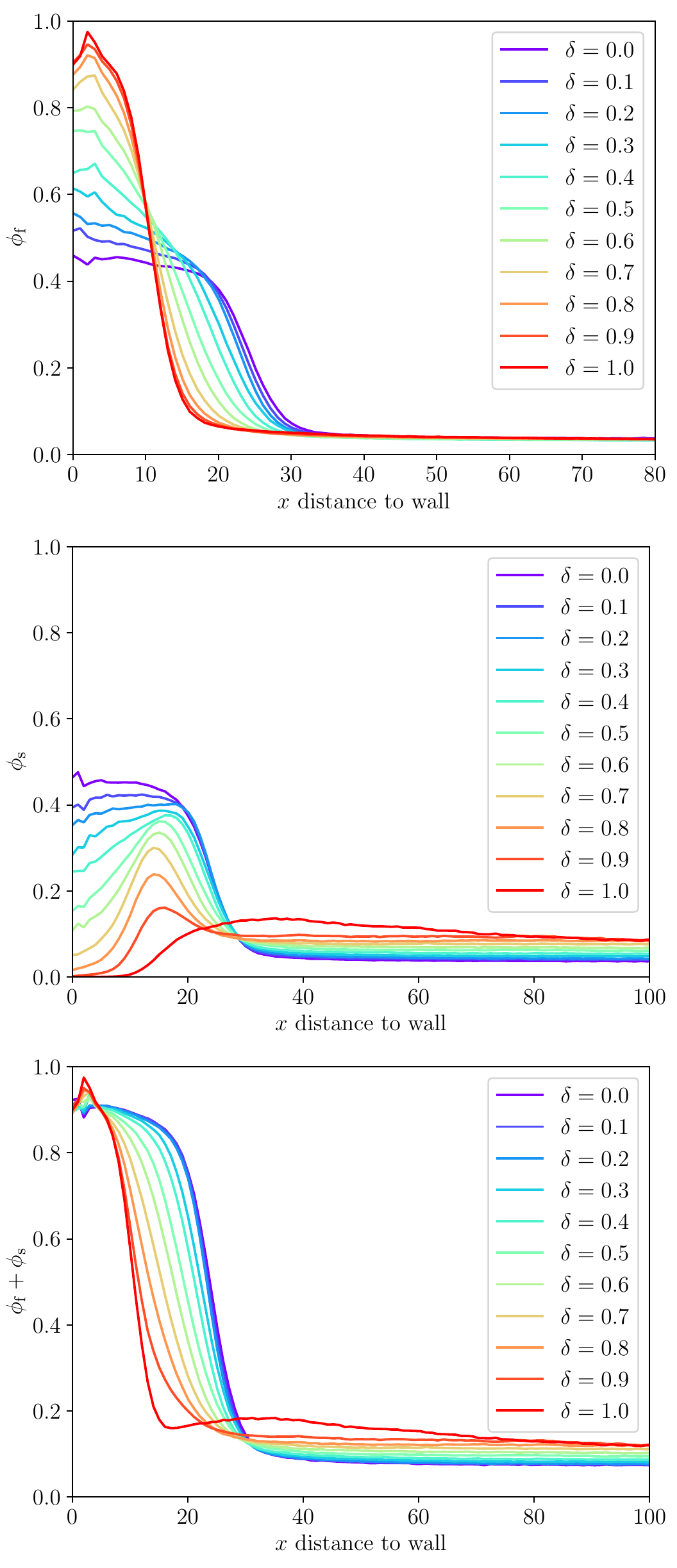}
		\caption{Stationary concentration profiles for slow [$\phi_{\rm s}(x)$] (top), fast [$\phi_{\rm s}(x)$] (medium), and all [$\phi_{\rm f}(x)+\phi_{\rm s}(x)$] (bottom) particles for various values of speed diversity degree $\delta$. The global area fraction is $\phi=0.18$.}
\label{fig.segregation_2}
\end{figure}

Finally, in Fig.~\ref{fig.segregation_2}, we present the stationary concentration profiles for slow, fast, and all  particles for various values of speed diversity degree $\delta$. Small oscillations arise from the fact that, similarly to the case of molecular fluids, near the walls, particles accumulate in a series of stable one-particle layers.

\section{Kinetic theory}
To understand the above results, we develop a simple kinetic theory that estimates the emission and absorption rates of fast and slow particles, $k_\text{out}^\text{(f/s)}$ and $k_\text{in}^\text{(f/s)}$, and thus the layer thickness and composition versus time. For that, we generalize a previous theory originally developed for systems without walls \cite{redner} to include mixtures (beyond the simpler approximation for mixtures in Ref.~\cite{kolb2020active}). 

Since the global density is low, we use an ideal gas approximation in the gas (see Fig.~\ref{fig.segregation_2}), i.e., particles there do not interact. The rate of absorption of particles by the layer, i.e., the incoming flux per unit length, is written as $k_{\rm in}^\text{(f/s)}=\frac{\rho_{\rm g}^\text{(f/s)}}{2\pi}\int_{-\pi/2}^{\pi/2}v^\text{(f/s)}\cos \theta d\theta=\frac{\rho_{\rm g}^\text{(f/s)}v^\text{(f/s)}}{\pi}$, where we integrate $v_x$ over random orientations leading to the particle entering the layer (on the right without loss of generality) weighted by the distribution $\rho_g^\text{(f/s)}/2\pi$ where $\rho_g^\text{(f/s)}$ are the gas number densities in contact with the layer. %
For a time $t^*_\text{(f/s)}$, $\rho_g^\text{(f/s)}$ is approximately constant and equal to the initial density as the front of non-interacting gas particles arrives at the layer. Only later, once the  ``cleaning signal'' mentioned before has overcome the entire system, $\rho_g^\text{(f/s)}$ evolves into the current bulk gas density obtained from the absorption-evaporation balance. We estimate $t^*_\text{(f/s)}=L_x/\langle v_{x}^{(\rm f/s)}\rangle$. Finally, $\rho_g^\text{(f/s)}$ are assumed to change abruptly at $t^*_\text{(f/s)}$ between their two values. 
This approximation, which leads to abrupt changes in layer growth rate at two instants, is more accurate for fast particles (see Fig.~\ref{orievol}-left), as the crossover time is smaller and the rotational diffusion has not significantly acted yet; for slow particles, the transition is smoother (see Fig.~\ref{orievol}-right).

\begin{figure}[htbp]
		\includegraphics[width=\linewidth]{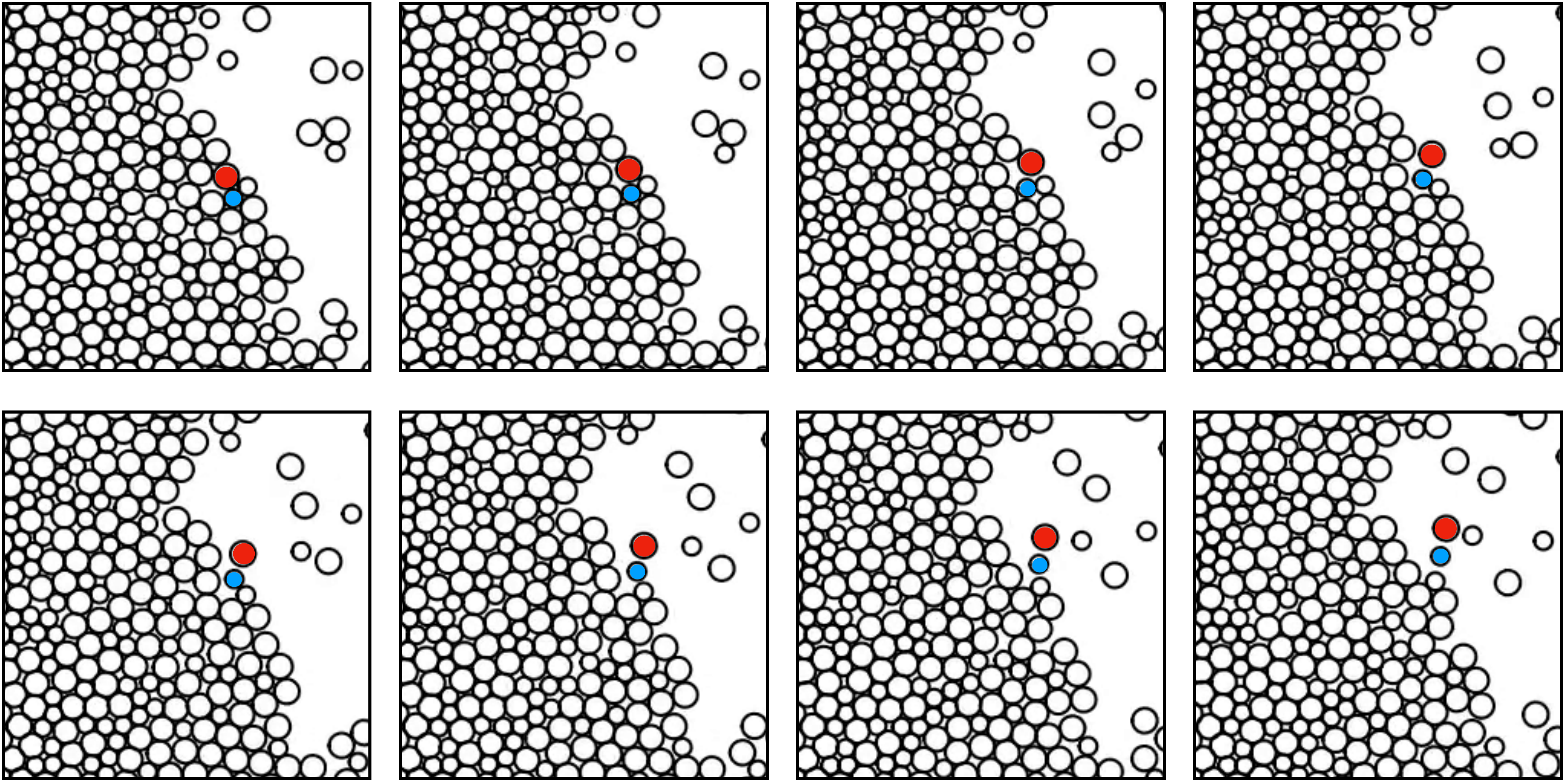}
		\caption{Sequence of snapshots (from left to right and top to bottom), showing an avalanche event for $\delta=0$. The first escaping particle is shown in red, followed by a small avalanche of one particle (in light blue).}
	\label{fig.avalanche}
\end{figure}

The SS $k_{\rm out}$ is calculated by solving the diffusion equation in angular space for $P$, the distribution of orientations at the interface, i.e., $\partial_t P(\theta,t)=\eta \partial_{\theta}^2P(\theta,t)$
with absorbing boundaries at $\pm\pi/2$ and initial condition given by the distribution of incident particles, i.e., $P(\pm\pi/2,t)=0$ and $P(\theta,0)=\cos{\theta}/2$ (as particles with $|\theta|\geq\pi/2$ cannot reach the wall and those with adequate $\theta$ will hit it with probability proportional to the $x$-axis velocity, normalized by integrating between $\pm\pi/2$). The solution is $P(\theta,t)=e^{-\eta t}\cos \theta/2$.
For average diameter $\sigma$ and identical speeds, one can write
$k_{\rm out}\equiv -\frac{\dot{N}_{\rm interface} }{\sigma N_{\rm interface}}=\frac{\eta}{\sigma}\rightarrow \frac{\kappa \eta}{\sigma}$
where $N_{\rm interface}=\int_{-\pi/2}^{\pi/2} P(\theta,t)$ is the number of particles at the interface and the dot is the time derivative. The result is corrected by a factor $\kappa$: when a particle escapes, some inner particles pointing towards the gas follow it in an avalanche-like effect (see Fig.~\ref{fig.avalanche} for an example of such phenomenon). In the SS, the average number of particles leaving the layer per escape event is denoted $\kappa=1+\kappa_{\rm excess}$. The value of $\kappa_{\rm excess}$ is treated as a fitting parameter (Ref.~\cite{redner} found that $\kappa_{\rm excess}\approx 3.5$ works well for all studied $v$ and $\phi$ in one-component systems without walls; in 1D, $\kappa_{\rm excess}=1$ \cite{soto2014run}). 
However, since at early stages particles in the layer are highly oriented towards the wall (see Fig.~\ref{orievol}), avalanche effects become strong only after $\tau$. Before that, once a particle escapes, other particles are likely to be still pointing towards the wall and therefore will not escape. This is incorporated by considering that $\kappa$ is time-dependent: $\kappa(t)=1+\kappa_{\rm excess}(1-e^{-\eta t})$, meaning that avalanche events occur with probability $(1-e^{-\eta t})$ as particles start to rotate away from the wall.
Finally, with speed diversity, one has
\begin{equation}
    k_\text{out}^\text{(f/s)}=\cfrac{N_\ell^\text{(f/s)}}{N_\ell^\text{(f)}+N_\ell^\text{(s)}}\cfrac{\kappa(t) \eta}{\sigma}, \label{eq.kout}
    \end{equation}
where $N^{\rm (f/s)}_{\rm \ell}$ is the number of particles of each type in the layer. Crucially, the factor $N_\ell^\text{(f/s)}/\left(N_\ell^\text{(f)}+N_\ell^\text{(s)}\right)$, which states that particle emission is taken as proportional to the fraction of particles of each type in the layer, nonlinearly couples the occupations of both types.

\begin{figure}[htbp]
	\centering
	\includegraphics[width=.5\textwidth]{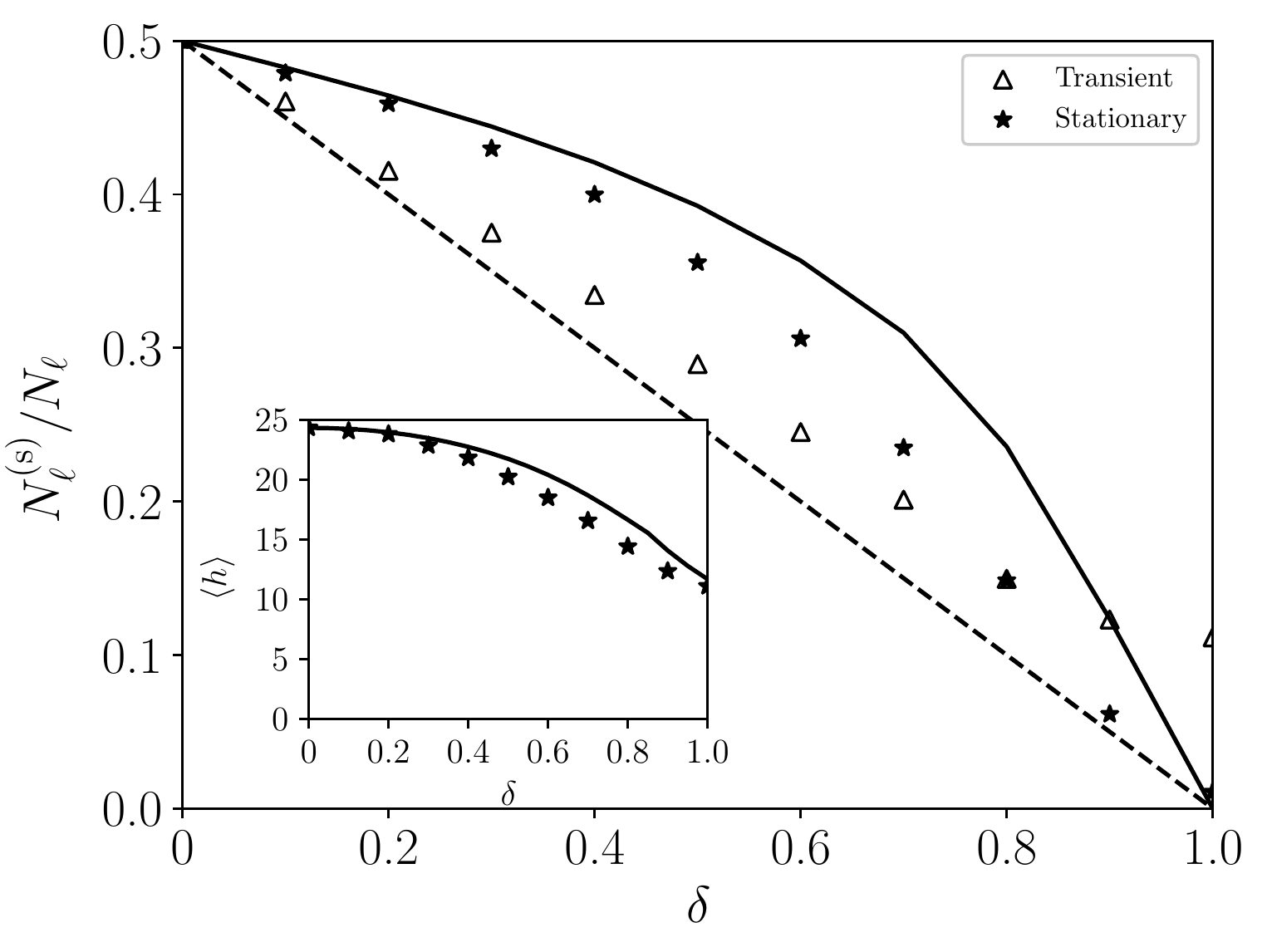}
	\caption{
	Ratio between the number of slow particles and the total number of particles in the layer for $\phi=0.18$ as a function of $\delta$. Dashed and solid lines show the theory for the transient and for the SS, respectively. Triangles and stars show simulation results for the transient and for the SS, respectively.	
	Inset: Stationary mean layer thickness. Stars are simulation and the solid line is the theory. 
	}
	\label{h_inf}
\end{figure}

The evolution of the parameters involved in the absorption and emission rates provides the layer thickness and composition at any time via $dN_\ell^{\rm (f/s)}/dt=\left(k_{\rm in}^{\rm (f/s)}-k_{\rm out}^{\rm (f/s)}\right)L_y$. Assuming particle conservation and that the layer is rectangular and close-packed with the hard-disk occupied area fraction $\phi_{\rm cp}=\pi/(2\sqrt{3})$ (as observed in simulations; see Fig.~\ref{fig.segregation_2}), we obtained a theory for $\langle h(t) \rangle$.
A good agreement occurs for $\delta$ near $0$ and near $1$---see left side of inset of Fig.~\ref{init}(b). For intermediate $\delta$, the value of $t^*_{\rm s}$ grows larger than the diffusion time and the theory becomes less good for intermediate times---see right side of inset of Fig.~\ref{init}(b). 
Also, the theory predicts that the initial deposition rate is independent of $\delta$, $d{N_\ell}/dt=L_y(\rho_0v_0/\pi-\eta/\sigma)$, in agreement with the simulations [Fig.~\ref{init}(b)].
Notably, the overshoot in $\langle h(t) \rangle$ is well captured, which is not the case if either the effect of $t^*$ or the relaxation of $\kappa$ are not included in the model. The layer composition evolution is also well captured (Fig.~\ref{Composition}-bottom), showing also the two stages found in the simulations. For $\delta=1$, the theory predicts no slow particles are in the layer as they are nonmotile; however, in simulations the transient concentration of slow particles is finite, with an ulterior elimination of them. This difference, also present for $\delta=0.9$, is due to an induced accumulation of slow particles pushed by fast ones, an effect that is neglected by the ideal gas assumption in the gas.
Figure~\ref{h_inf} compares theory and simulation for the fraction of slow particles in the layer. In the inset, this comparison is shown for the SS layer thickness, with very good agreement.

\section{Conclusions}
Self-propelled Brownian particles under repulsive interactions spontaneously exhibit complete wetting layer formation in the presence of a flat wall due to persistent motion. With simulations and a theory with speed diversity, we calculate and explain the wetting layer composition and thickness. We 
reveal a two-stage evolution for the layer composition and a transient overshoot for the layer thickness, explained only when the theory considers delayed avalanche-like emissions outwards and a transient front of particles moving towards the walls. 

An implicit assumption of the theory is that no segregation develops inside the layers [mean-field approximation in Eq.~\eqref{eq.kout}]. However, Figs.~\ref{init}(a) and \ref{fig.segregation_2} indicate that spatial segregation does exist, with fast particles closer to the wall; a more detailed analysis is beyond our scope here.
Note that in Ref.~\cite{kolb2020active}, for a much denser case showing bulk MIPS ($\phi=0.6$), the opposite is seen: faster particles accumulate at cluster boundaries.

Since bacteria located inside thick layers may be protected, our work shows how biological variability of motility properties can play a central role in determining the survivability of microorganisms. More broadly, our results provide important insights into the behavior of active matter such as the origin of swim pressure overshoots previously seen in confined systems \cite{patch2017kinetics}. Furthermore, our framework can be adapted to study bacterial types competing to colonize niches in confined systems \cite{hibbing2010bacterial} as well as the puzzling formation of multi-cellular aggregates such as ameboid slime mold, where slower cells hijack the motion of faster cells to move further and spread their spores at low energy cost~\cite{miele2021aggregative}.


\section*{Acknowledgments} MR-V and RS are supported by Fondecyt Grant No.~1220536 and ANID -- Millennium Science Initiative Program -- NCN19\_170D, Chile. PdC is supported by grant \#2021/10139-2, São Paulo Research Foundation (FAPESP), Brazil.


%


\end{document}